\documentclass[preprint]{aastex}

\newcommand{\beq}{\begin{equation}}
\newcommand{\eeq}{\end{equation}}
\newcommand{\be}{\begin{eqnarray}}
\newcommand{\ee}{\end{eqnarray}}
\newcommand{\rar}{\rightarrow}
\newcommand{\lrar}{\leftrightarrow}
\newcommand{\fg}{f_\gamma}

\shorttitle{Spectral distortion of CMBR}
\shortauthors{Dolgov, Hansen, Semikoz, Pastor}

\begin{document}


\title{Spectral distortion of cosmic background radiation by\\
scattering on hot electrons. Exact calculations.}


\author{A.D. Dolgov\altaffilmark{1}, S.H. Hansen}
\affil{INFN section of Ferrara\\Via del Paradiso 12, 44100 Ferrara, Italy}
\email{dolgov@fe.infn.it, sthansen@fe.infn.it}

\author{S. Pastor\altaffilmark{2}, D.V. Semikoz\altaffilmark{3}}
\affil{Max-Planck-Institut f\"ur Physik (Werner-Heisenberg-Institut)\\
F\"ohringer Ring 6, 80805 M\"unchen, Germany}
\email{pastor@mppmu.mpg.de, semikoz@mppmu.mpg.de}


\altaffiltext{1}{ITEP, Bol. Cheremushkinskaya 25, Moscow 117259, Russia}
\altaffiltext{2}{SISSA--ISAS and INFN section of Trieste,
Via Beirut 2-4, 34014 Trieste, Italy}
\altaffiltext{3}{Institute of Nuclear Research of the Russian Academy
of Sciences, 60th October Anniversary Prospect 7a, Moscow 117312,
Russia}


\begin{abstract}
The spectral distortion of the cosmic background radiation produced by
the inverse Compton scattering on hot electrons in clusters of
galaxies (thermal Sunyaev--Zel'dovich effect) is calculated for
arbitrary optical depth and electron temperature. The distortion is
found by a numerical solution of the exact Boltzmann equation for the
photon distribution function. In the limit of small optical depth and
low electron temperature our results confirm the previous analyses. In
the opposite limits, our method is the only one that permits to make
accurate calculations.
\end{abstract}


\keywords{cosmic microwave background --- cosmology: theory ---
galaxies: clusters: general --- methods: numerical --- scattering}


\section{Introduction}

The frequency spectrum of the Cosmic Microwave Background Radiation
(CMBR) is known to have the equilibrium black body form (in natural
units in which $\hbar=c=k_B=1$)
\beq
f_0 (p_\gamma) = \left[ \exp (p_\gamma /T_\gamma) -1\right]^{-1} \, ,
\label{f0}
\eeq
with the temperature $T_\gamma = 2.725 \pm 0.002$ K \citep{cobelast}.
No deviation from this perfect Planck distribution is observed with
the accuracy of $10^{-4}$.  These data present a strong evidence in
support of Big-Bang cosmology. It is established that the cosmic
plasma in the early universe was in a thermal equilibrium state and
the spectrum remained practically undisturbed to the present
epoch. Nevertheless, small deviations from the perfect Planck spectrum
are possible, and they could provide interesting information about
physical processes in the early universe that might take place at
relatively small red-shifts, $z \leq 10^7$. At higher red-shifts all
distortions of the thermal equilibrium spectrum would be efficiently
smoothed down by the Compton scattering, $\gamma +e \lrar \gamma +e$
and by the inelastic photon producing reactions, double Compton,
$\gamma + e \lrar 2\gamma +e$ or Bremsstrahlung, $e + A \lrar e + A +
\gamma$. The first elastic process restores kinetic equilibrium,
i.e. it forces the photon distribution function, $f_\gamma$, to take
the Bose-Einstein form with a possible non-zero chemical potential,
$\mu$, while the other two reactions push $\mu$ down to zero.
 
There are several possible sources and mechanisms that could give rise
to spectral distortions of the CMBR both in the early and the present
day universe. In the early universe there could be electromagnetic
decays of long-lived particles with life-time larger than $\sim 100$
sec, in particular, $\nu_H \rar \nu_L + \gamma$. A study of a possible
distortion of the CMBR spectrum permits to obtain strong bounds on the
probability of such decays (for a review, see \citet{cmb-limits}). In
the present-day universe a very interesting spectral distortion can be
induced by the scattering of cosmic background radiation on hot
electrons in galactic clusters, known as the Sunyaev-Zel'dovich (SZ)
effect~\citep{zs1,zs2,zs3,zs4}. Observations of such a distortion,
combined with those of thermal X-ray emission of the cluster gas,
could help to extract important astrophysical information, in
particular to measure the Hubble constant, or to study the evolution
of clusters (for reviews see \citet{review-zs1,review-zs2,revReph,
Birkinshaw99}, and the updates in \citet{3Kb} and \citet{3Kr}).

Compton scattering conserves the number of photons, which implies that
the SZ effect produces a systematic shift of photons from the
low-energy part to the Wien side of the Planckian CMBR spectrum. To
calculate the corresponding distortion of the spectral distribution
one should use the Boltzmann kinetic equation for the distribution
function of photons, $f_\gamma$,
\beq
{d f_\gamma \over dt} = I_{coll} +S \, ,
\label{kineq}
\eeq
with properly taken collision integral $I_{coll}$ and a possible
source term, $S$.

In the early universe case one has to solve the system of such two
coupled equations for the photon and electron/positron distributions. In
the lowest order in the fine-structure constant, $\alpha \simeq
1/137$, only Compton and Coulomb scattering need to be taken into
account.  They have cross-sections of the order $\alpha^2$, while
inelastic reactions contribute at most to the order
$\alpha^3$. Because of technical difficulties this problem was treated
only approximately especially for relativistic electrons. In the
non-relativistic case the problem can be reduced to the partial
differential equation describing diffusion in photon momentum
space. For the particular case of Compton scattering such a reduction
was done in 1957 by Kompaneets \citep{kompaneets57} and for inelastic
processes (double Compton scattering and Bremsstrahlung) by Lightman
\citep{lightman81}. Solutions to these equations in cosmological
situations were analyzed in \citet{bernstein90} and \citet{hu93}.

In the case of scattering of the CMB radiation on hot electrons in
galactic clusters (thermal SZ effect) the Kompaneets equation has the
form
\beq
{\partial f_\gamma (X,t) \over \partial t} =
n_e \sigma_T\, {T_e \over m}\, {\partial \over X^2 \partial X}
\left[ X^4 \left( {\partial \fg \over \partial X} +\frac{T_\gamma}{T_e}\fg(1 +\fg)
\right)\right] \, ,
\label{komp}
\eeq
where $T_e$ and $T_\gamma$ are the electron and photon temperatures, 
$m$ is the electron mass,
\be
X=\frac{p_\gamma}{T_\gamma}
\label{defX}
\ee 
is the dimensionless photon momentum, $n_e$ is the
electron number density, and $\sigma_T = 8\pi\alpha^2 /3m^2 =
6.65\times 10^{-25}\, {\rm cm}^2$ is the Thomson cross-section.
However, the electrons in the clusters are hot, with temperatures that
can be larger than $15$ keV, which means that the energy change of
Compton scattered photons is not small enough to be accurately
described by the Kompaneets equation.

There are basically two different approaches in the literature to
extend the validity of this equation. In ref.~\citet{Rephaeli95} the
relativistic form of the Maxwell velocity distribution of electrons is
used and the frequency redistribution through Compton scattering is
calculated (see also \citet{Wright79,Yankovitch97,Sazonov98,Molnar99});
this method is also known as the Radiative Transfer Approach. On the
other hand, one can obtain a relativistic generalization of the
Kompaneets equation, by expanding in series of the parameter $\theta_e
\equiv T_e/m$.  Low orders of this expansion were considered in the
papers \citet{Stebbins97} and \citet{Challinor98}, while 
\citet{ItohI} took into account relativistic corrections up to
${\cal O}(\theta_e^5)$. These two different methods are essentially
consistent.

It has, however, been pointed out \citep{Challinor98} that these
results are not guaranteed to be accurate because the series
approximation to the solution of the Boltzmann kinetic equation
involves expansions in parameters which are not  small. In
particular it was recognized that the convergence is slow and possibly
even asymptotic (for instance, to calculate the crossover frequency,
where the thermal distortion vanishes, it is better to use a linear
approximation in $\theta_e$ than the expansion up to ${\cal O}
(\theta_e^3)$ or ${\cal O} (\theta_e^5)$).  
In view of that it is desirable to solve the original kinetic equation
(\ref{kineq}) directly without any specific approximation. This
problem was partly addressed in \citet{ItohI,ItohIV} where the
collision integral was numerically calculated for the unperturbed
photon distribution eq.~(\ref{f0}).  This integral determines the
first time derivative of the distribution function $f_\gamma (X, 0)$
and in the limit of a small optical depth,
\be
\tau = \int dl~n_e \sigma_T \ll 1 \, ,
\label{tau}
\ee
(the integral is taken along the line of sight through the cluster)
gives the solution, $\fg (X,\tau) \approx f_0 + \tau \,I_{coll} [\fg
(X,0)]$.  In a subsequent paper \citep{ItohV}, the second order
corrections in $\tau$ were calculated as a series in $\theta_e$. The
results show a good agreement with solutions of the generalized
Kompaneets equation in the limit of small optical depth and low
electron temperatures, $\theta_e =T_e /m \la 0.03$ (i.e. $T_e \la 15$
keV).

In this paper we present an accurate numerical solution of the exact
Boltzmann kinetic equation for arbitrary optical depth and electron
temperature. In the limit of small $\tau$ and $\theta_e$ our results
confirm those of the previous papers, and in particular of
\citet{ItohI,ItohV}.  However, the method presented here permits to
solve the equation precisely for an arbitrary optical depth and any
isotropic electron distribution function. We use essentially the same
method as the one we have developed for the calculations of the
spectral distortion of light \citep{dolgov97,dolgov99a} or heavy
\citep{dolgov98,dolgov99b} neutrinos at Big Bang Nucleosynthesis.  An
essential point of the calculations is an analytical reduction of the
exact collision integral down to two dimensions
\citep{TkachevI,TkachevII}. In the case of Compton scattering the
problem is much more complicated because the squared amplitude of this
process is not a simple polynomial function of particle momenta as was
the case for weak interaction in the low energy limit.

A reduction of the collision integral down to one dimension in the
direct reaction term and down to two dimensions in the inverse
reaction term was done in \cite{Poutanen96} for the case of
Boltzmann statistics (see also the review \cite{Revpou} for a
discussion and relevant references). The method of these works is
somewhat different from ours because we specially use the procedure of
the integration suitable for the quantum statistics case when the
direct reaction term (as well as the inverse one) contains the product
of the distribution functions not only in the initial state but also
in the final state, $f_1 f_2 (1+ f_3)(1-f_4)$ (see eq.~\ref{Ff}), that
makes the reduction down to one dimension impossible in principle if
all the functions are considered as unknown.

The paper is organized as follows. In the next rather technical
section we present the reduction of the 9-dimensional collision
integral down to a 2-dimensional one. In section 3 the numerical
solution of the integro-differential Boltzmann equation is
described. In section 4 our results are presented and
discussed. Finally in section 5 we give the conclusions.

\section{The Boltzmann equation}

We consider the Boltzmann equation (\ref{kineq}) for the distribution
function of photons, $f_\gamma$, taking into account only Compton
scattering process,
\beq
\gamma(P_1) + e(P_2) \leftrightarrow \gamma(P_3) + e(P_4) \, .
\label{process}
\eeq
where $P_i = (E_i,{\mathbf p_i})$ are the particle 4-momenta.  In this
case the source term in the kinetic equation is absent, $S=0$, and we
have \be {d f_\gamma \over dt} = I_{coll} \, ,
\label{kin2}
\ee 
where the collision integral $I_{coll}$ takes the form
\beq
I_{coll} = \frac{1}{2E_1}~\int
\prod_{i=2}^4\left (\frac{d^3p_i}{(2\pi)^3 2E_i}\right ) 
~(2\pi)^4\delta^4(P_1+P_2-P_3-P_4)~|M|^2~F(1,2,3,4) \, ,
\label{icoll}
\eeq
with the statistical factor, $F$, given by
\beq
F = f_\gamma(p_3) f_e(p_4) [1+f_\gamma(p_1)][1-f_e(p_2)] -
f_\gamma(p_1) f_e(p_2) [1+f_\gamma(p_3)][1-f_e(p_4)]  \, ,
\label{Ff}
\eeq
and $|M|^2$ is the matrix element squared of the process
in eq.~(\ref{process})
\be
|M|^2 &=& 4e^4\left\{ m^4
\left[\frac{1}{(P_1\cdot P_2)^2}
-\frac{2}{(P_1\cdot P_2)(P_1\cdot P_4)}
+\frac{1}{(P_1\cdot P_4)^2}
\right] \right.
\nonumber \\
 &+& \left. 2 m^2\left[ \frac{1}{(P_1\cdot P_2)}- \frac{1}
{(P_1\cdot P_4)}\right]
+ \frac{(P_1\cdot P_4)}{(P_1\cdot P_2)} 
+ \frac{(P_1\cdot P_2)}{(P_1\cdot P_4)} \right\} \, ,
\label{matrix}
\ee
where $m$ is the electron mass.

The collision term in eq.~(\ref{icoll}) is in principle a
9-dimensional phase space integral, that can be analytically
reduced to a sum of two-dimensional integrals. When one defines the
terms $I_k$ for $k=1, \ldots , 7$, corresponding to the 7 terms in
eq.~(\ref{matrix}), each of them reduces to
\beq
I_k (p_1) = \frac{\alpha^2}{2\pi p_1^2} \int_0^\infty~
\frac{p_2dp_2}{E_2}~\int_0^{{\cal S}}~dp_3~F~ J_k(p_1,p_2,p_3)~,
\label{2-dim}
\eeq
where $\alpha \simeq 1/137$ is the fine-structure constant and the
maximum value $S$ for the momentum $p_3$ is discussed in appendix
\ref{integration}. The expressions for the integrands $J_k$ are listed
in appendix \ref{reduction}. Each of the two-dimensional integrals
in eq.~(\ref{2-dim}) can be found numerically, which permits to
evaluate $I_{coll}$ in eq.~(\ref{icoll}).  The spectral distortion of
$f_\gamma$ is then found from the evolution of the Boltzmann equation,
eq.~(\ref{kin2}) (see the next section for the computational details).
An important issue, which we have taken into account, is the reduction
of the integration region in eq.~(\ref{2-dim}). When one imposes
conservation of 3-momenta, and in particular the fact that none of the
momenta can be larger than the sum of the others, i.e.~$p_i \leq p_a +
p_b +p_c$, the integration region can be divided into several
sub-regions where the functions $J_k$ take different values. See the
Appendices for the technical details.

\section{Numerical solution of the Boltzmann equation}

We have found the spectral distortion of $f_\gamma$ produced by the SZ
effect from the numerical solution of the Boltzmann equation (\ref{kin2}). 

We measure all momenta in units of photon temperature.  Accordingly,
we use the dimensionless photon frequency $X$ from eq.~(\ref{defX})
and measure time in units of the optical depth, $\tau$, given by
eq.~(\ref{tau}).  In order to calculate the integral over electron
momenta we introduce the electron temperature as
\be
X_e = \frac{p_e}{T_\gamma} = R_T \frac{p_e}{T_e}~,
\ee
where 
\be
R_T = \frac{T_e}{T_\gamma}~,
\label{rel_T} 
\ee  
is the ratio of electron to photon temperatures.
We assume that electrons are always in equilibrium with the distribution
\be
f_e  = \left(e^{(E_e - \mu_e)/T_e} +1 \right)^{-1}~,
\label{f_e}
\ee where $E_e = \sqrt{m^2+p_e^2}$ is the electron energy and $\mu_e$
is their chemical potential. In the present paper we make the
simplifying assumption of Boltzmann statistics\footnote{Note that our
method is valid for any isotropic $f_e$, not necessarily in
equilibrium, but here we restrict ourselves to the case of a Boltzmann
equilibrium distribution in order to investigate the thermal SZ
effect.} for electrons ($f_e \ll 1$) and
express $\mu_e$ through the number density of electrons.

The integration over electron momenta in the two-dimensional integral
in eq.~(\ref{2-dim}) should formally be taken from zero to infinity,
but we integrate only to $p_e^{\rm max}$.  We found that for small
$T_e < 10 ~{\rm keV}$ it is enough to take $p_e^{\rm max} = m$, and
for larger electron temperature $T_e \sim 50 ~{\rm keV}$ integration
till $p_e^{\rm max} = 3 m$ is sufficient.

For the dimensionless photon frequency, $X$, we introduce a grid of
$N_{\rm grid}$ points, logarithmically spaced in the interval $X_{\rm
low} \le X \le X_{\rm hi}$.  We use $X_{\rm low}=10^{-4}$, which is
small enough to maintain a sufficiently high
precision of the calculations. $X_{\rm hi}$
actually depends both on the electron temperature $T_e$ and the final
optical depth $\tau_{\rm fin}$.  For small electron temperatures $T_e
< 10 ~{\rm keV}$ it is enough to take $X_{\rm hi}=100$. For higher
temperatures and large optical depth $\tau > 1$ we need to use much
higher values of $X_{\rm hi}$, because photons 
multiple-scatter into large
momentum modes.  For example, for calculation with $T_e=20 ~{\rm keV}$
and $\tau_{\rm fin}= 20$ (see the next section) we use $X_{\rm hi} =
10000$.  As for the number of points in the grid, $N_{\rm grid}$, we
always checked that it is large enough to maintain the desired
precision.  Typically 1600 points grid was good enough for our
purposes.

As initial condition for eq.~(\ref{kin2}) we used equilibrium photon
distribution function eq.~(\ref{f0}), which in terms of $X$ has the
simple form
\be
f_0(X) = \frac{1}{e^{X}-1}~. 
\label{f_eq}
\ee 

For calculation of the two-dimensional integrals in eq.~(\ref{2-dim})
we use the Gaussian quadratures method, see \citet{numrec}.  For the
evolution in "time", $\tau$, we use the simple Euler method, which is
precise enough for the present calculations.  Finally, for calculation
of the photon distribution function in points between the grid points
we used interpolation not in $f_\gamma$ but in $f_\gamma - f_0$ to 
perform 2D-integration. This trick strongly reduces
numerical errors, coming from interpolation in the
regions where the distribution of photons is close to the equilibrium
one (i.e. for $\tau < 1$).

Additional difficulties in the numerical calculations come from the
region of small photon momenta, $X < 0.1$. In this region some parts
of the integrands in eq.~(\ref{2-dim}) behave like 
$1/X^5$, while the complete collision
integral is proportional to $1/X$ for small $X$ (due to the
singularity $1/X$ in the 
distribution function eq.~(\ref{f_eq})).  This leads to large
numerical errors for small momenta. On the other hand, asymptotically
for small $X$ each part of the collision integral eq.~(\ref{2-dim})
has either a constant value or $1/X$-behavior. Using this fact, we
calculate the integrals numerically for grid points with
momenta $X > X_{\rm small}$ and extrapolate into the region of smaller
$X$. In the calculations we use $ 0.01\le X_{\rm small} \le 0.1$.

Different parts of the integration region contribute differently to
the collision integral in eq.~(\ref{2-dim}) (see appendix
\ref{integration} for the definitions of the regions). The main
contribution comes from the large regions $C_2$ and $D_2$. The regions
$C_1$ and $D_1$ can be taken into account, but they contribute on the
level of 1\% in the region of small $X < 0.1$ and could be neglected
for larger $X$. The main numerical error comes from a small part of
the integration regions $C$ and $D$ around the line ${\cal Q}^2=0$.
These regions give numerical errors up to 10\% of the value of the
collision integral for small photons momenta. The nature of this
small-momentum error was discussed above. The regions $A$ and $B$ give
corrections to the collision integrals in the sixth digit and could be
neglected.

The ratio of the electron and photon temperatures,
$R_T$~(\ref{rel_T}), is very large in hot clusters. The CMB
temperature is $T_\gamma \sim 10^{-4}$ eV, while the electron
temperature is $T_e \sim 10 ~{\rm keV}$. Since both photon and
electron momenta are involved in the same expressions in the collision
integral we may lose precision by many orders of magnitude if we keep
realistic values of $R_T \sim 10^{8}$. The energy conservation implies
$p_3 = p_1 + E_2 -E_4$, and since $E_2 \sim E_4 \sim m = 511 ~{\rm
keV}$, the difference should cancel up to 10 digits in order to give
numbers of the order $p_1 \sim p_3 \sim 10^{-4}~{\rm eV}$, and even
more precisely for low photon momenta, $X<1$. This numerical problem
is solved in the following way. We calculate the collision integral
for much higher photon temperatures $T_\gamma \sim 1$ keV, and then
diminish the temperature until the collision integral stops varying
with a further temperature decrease.  We assume that at this value of
$T_\gamma^{\rm min}$ the collision integral reaches its asymptotical
value and therefore reproduces the result for $T_\gamma \sim
10^{-4}~{\rm eV}$.  We found that already at $T_\gamma = 50$ eV the
collision integral reaches asymptotical values. In most calculations
we used $T_\gamma^{\rm min} = 10$ eV. Let us note, that already for
$T_\gamma = 1$ eV the numerical errors discussed in this paragraph
become important.

In conclusion, we control all numerical errors in the solution of
the kinetic equation~(\ref{kin2}), and we also check the result 
testing the conservation of the photon number (see the 
end of section~\ref{sec:res}).

\section{Results}
\label{sec:res}
The distortion of the photon spectrum is usually presented in 
the form
\be
\Delta I(X,\tau) = X^3 (f_\gamma(X,\tau)-f_0(X))~, 
\label{distor}
\ee
where $f_\gamma(X,\tau)$ is the photon distribution function calculated 
from the kinetic equation~(\ref{kineq}) and $f_0(X)$ is the initial
equilibrium photon distribution (\ref{f0}).

Instead of solving the kinetic equation directly, one could for a small
optical depth, $\tau \ll 1$, make the Taylor expansion in $\tau$ of the
collision integral
\be
I_{coll}(X,\tau) \approx I_{coll}(X,0) + \tau \frac{\partial
I_{coll}}{\partial \tau}(X,0) + {\cal O}(\tau^2) \, .
\label{I_approx}
\ee
Then the distribution function $f_\gamma(X,\tau)$ can be trivially found from  
eq.~(\ref{kin2})
\be
f_\gamma(X,\tau) \approx f_0(X) + \tau I_{coll}(X,0) +
\frac{\tau^2}{2} \frac{\partial I_{coll}}{\partial \tau}(X,0) + {\cal
O}(\tau^3) \, ,
\label{f_approx}
\ee  
and the spectral distortion takes the form 
\be
\Delta I(X,\tau) \approx \Delta I_1(X,\tau) + \Delta I_2(X,\tau) + {\cal O}(\tau^3)~, 
\label{distor1a}
\ee
where
\be
\Delta I_1(X,\tau) = \tau X^3 I_{coll}(X,0)~~~,~~~
\Delta I_2(X,\tau) = \frac{\tau^2}{2}  X^3 \frac{\partial I_{coll}}{\partial \tau}(X,0)
\label{distor2}
\ee

The term $\Delta I_1$ in eq.~(\ref{distor2}) corresponds to the single
photon scattering.  It requires only calculation of the collision
integral with equilibrium distribution functions $f_0(X)$.  Our
calculation of the collision integral in this case agrees perfectly
with the results obtained in \citet{ItohI,ItohIV}, as can be seen from
fig.~\ref{fig:I1}. In this figure we compare our numerical results for
$T_e=10$ keV and $T_e=20$ keV with analytical fits from
\citet{ItohIV}.

The term $\Delta I_2$ in eq.~(\ref{distor2}) is proportional to
$\tau^2$, and it corresponds to the double photon scattering
contributing to the spectral distortion.  In order to calculate this
term one can do the following: calculate $I_{coll}(X,0)$, substitute
the first two terms of the photon distribution function
eq.~(\ref{f_approx}) back into the collision integral and calculate
all terms of order $\tau$. These terms correspond to the second term
in eq.~(\ref{I_approx}), which is responsible for the double photon
scattering.  In the paper~\citet{ItohV} this second order contribution
to the spectral distortion was found approximately, up to the terms of
the order $\theta_e^2$.  By numerical integration of eq.~(\ref{kin2})
we have found the two-scattering contribution exactly from
$I_{coll}(X,\tau)$.

In fig.~\ref{fig:I2} the double-scattering contribution into the
spectral distortion as a function of dimensionless photon frequency
$X$ is presented for $T_e=10$ keV. The solid curve in
fig.~\ref{fig:I2} is our exact result (up to small numerical errors).
One could compare it with the series expansion in powers of
$\theta_e=T_e/m$ from \citet{ItohV}, which are shown by the
dashed, short dashed and dotted lines (the definitions of functions
$Z_i$ are given in \citet{ItohV}).  As we can see from
fig.~\ref{fig:I2}, this expansion is in a good agreement with our
result for relatively small $X < 4$ and disagrees for a larger $X$.
It demonstrates again (as the first-order terms in $\tau$) a bad
convergence of the expansion in terms of $\theta_e$.

In figure~\ref{fig:T20y1} we present the results for the spectral
distortion of the CMB photons caused by the SZ effect for $T_e=20$ keV
and the optical depth equal to one, $\tau=1$. 
 The dotted curve
corresponds to the numerical solution
of eq.~(\ref{kin2}). 
The solid line represents the single scattering
contribution, namely the first term in eq.~(\ref{distor1a}).  The long
dashed line corresponds to the double scattering contribution, namely
the second term in eq.~(\ref{distor1a}).  The short dashed line is the
sum of these two contributions. One can see from fig.~\ref{fig:T20y1}
that even at optical depth $\tau=1$, the spectral distortion is
accurately described by the sum of single and double photon
scatterings.

It should be noted that our results have been obtained with an initial
isotropic photon distribution. This is an approximation of the real
radiation distribution inside the clusters of galaxies, which are of
finite extension. At the level of single scattering there is no
problem since the radiation is isotropic prior to convolving with the
Compton kernel. However the multiple scattering contributions to the
spectral distortion could be affected by the geometrical properties of
the cluster. For instance, \citet{Molnar99} used a Monte-Carlo method
to take into account finite optical depth and bulk motion in a
spherically symmetric uniform plasma. Therefore our results are exact
in the case of an infinite medium.

For $\tau \ll 1$, the
single scattering strongly dominates the spectral distortion, and in
order to estimate the contribution of the second-order term, one
should rescale the amplitude of the long dashed line in
fig.~\ref{fig:T20y1} by the factor $\tau$ and compare it with the
solid line at any photon frequency $X$. In a realistic case one has
$\tau\leq 0.01$, and the double scattering can be safely neglected
everywhere, except for a small region in photon frequency $X$ where
the single scattering contribution vanishes.

The magnitude of the double photon scattering is demonstrated in
figure \ref{fig:corr_Te} for the optical depth $\tau=0.01$.  The
crossover frequency $X_0$ (the photon frequency at which the
distortion caused by the thermal SZ effect vanishes) is plotted as a
function of the electron temperature (solid line). The regions around
this line show the double scattering contribution to the spectral
distortion. In the region between the dotted lines it is larger than
50\%, between the short dashed lines it is larger than 10\%, and
between the long dashed lines it is larger than 1\%.

When one calculates the distortion using the Kompaneets equation
(\ref{komp}), the cross\-over frequency does not depend on the
electron temperature, and it takes the constant value $X_0 \simeq
3.83$. The exact calculations show that the crossover frequency
depends both on the electron temperature, $\theta_e = T_e/m$ and the
optical depth $\tau$. For small optical depths, $0<\tau<0.05$, a
linear fit in $\tau$ is very accurate
\be
X_0 = \alpha(T_e) + \tau \, \beta(T_e) \, ,
\ee
and by doing such fits in the range, $0 < \tau < 0.05$, for many
different electron temperatures one can find $\alpha(T_e)$
and $\beta(T_e)$ as functions of temperature
\be
\alpha(T_e) \approx 3.830 (1+ 1.162 \, \theta_e - 0.8144 \, \theta_e^2) 
\, \, \, \, \, 
\mbox{and} \, \, \, \, \, 
\beta(T_e) \approx 3.021 \, \theta_e - 8.672 \, \theta^2_e \, ,
\ee
which fits better than $4 \times 10^{-4}$ for $0<T_e<50$ keV and
$0<\tau<0.05$.  The functional shape was chosen by simply looking at
the corresponding graphs. By considering the graphs for $\alpha(T_e)$
and $\beta(T_e)$, we have found no reason to include higher order
terms.
The fit parameters change only slightly for large optical depths,
$0.05 < \tau < 1$.  One can naturally perform similar fits for $X_{\rm
max}$ and $X_{\rm min}$, which are the dimensionless frequencies where
the spectral distortion is maximal or minimal. We find \be X_{\rm min}
&=& 2.265 \left(1 - 0.0927 \, \theta_e + 2.38 \, \theta^2_e \right) +
\tau \left( -0.00674 + 0.466 \, \theta_e \right)~, \\ X_{\rm max} &=&
6.511 \left( 1 + 2.41 \, \theta_e - 4.96 \, \theta^2_e \right) + \tau
\left( 0.0161 + 8.16 \, \theta_e - 35.9 \, \theta^2_e \right) .  \ee
We can compare these results with similar fits from the previous
papers. For the interesting range of temperatures, $0<T_e<50$ keV, and
for negligible optical depth, we are in good agreement with the
previous results~\citep{Birkinshaw99,Challinor98,ItohI}.  In
\citet{Molnar99} the fit in the optical depth is also made and for
small $\tau$ we are in good agreement. In this paper some of
the effects of the cluster geometry were considered, which
may explained why our results for larger $\tau$ 
are somewhat different.

Finally in figure \ref{fig:T20y20} the results for a large optical
depth, $\tau$, are presented. The Taylor expansion in terms of $\tau$,
eq.~(\ref{distor1a}), is useless for $\tau >1$, and the complete
numerical solution of eq.~(\ref{kin2}) should be done.  From figure
\ref{fig:T20y20} one can see that for $\tau > 1$ the spectral
distortion is dominated by multi-scattering contributions.  Indeed,
one and two-scattering contributions dominate the curves with
$\tau=0.01$ and $\tau=1$, and all distortion in these cases are in the
region $X<20$. Spectral distortion in higher modes come from
multi-scattering contribution, demanding much larger optical depth.

Asymptotically for $\tau \rar \infty$ the photon distribution should
reach the form
\be f_\gamma \rar \left[ \exp \left(\frac{E-\mu}{
T_e}\right) -1 \right]^{-1}~,
\label{distr-chempot} 
\ee 
with the chemical potential $\mu$ determined by the photon number
conservation, $\mu =T_e \, {\rm log}(\zeta (3) T_\gamma^3/T_e^3)
\approx -1$ MeV (when $T_\gamma \ll T_e$).  However, even for $\tau =
20$ the distribution is very far from the equilibrium one in
eq.~(\ref{distr-chempot}). A much larger $\tau$ is necessary to
equalize the temperatures of photons, $T_\gamma = 2\times 10^{-4}$ eV,
and electrons, $T_e \sim 20$ keV. We have estimated that one requires
$\tau \sim \sqrt{m/T_{e}} \ln ({\rho_{fin} /\rho_{in}})$, where
$\rho_{in}$ and $\rho_{fin}$ are the initial and final energy
densities of photons. We have also calculated the same quantity
numerically and found $\tau \sim 250$ for $T_e=20$ keV, in reasonable
agreement with this simple analytical estimate. As expected we observe
that the crossover frequency increases with rising $\tau$, and
asymptotically it should tend to infinity.

To check the precision of the calculations we test if the number
density of the photons is conserved, as it should be in elastic
Compton scattering.  For $T_e=20$ keV and 1600 point grid in photon
frequency the relative non-conservation of the photon number, $\Delta
N_\gamma /N_\gamma$, is approximately $10^{-9}$ for $\tau=0.01$ and is
smaller than $10^{-5}$ for any $\tau<20$.

\section{Conclusion}
We have numerically calculated the spectral distortion of the cosmic
background radiation produced by Compton scattering on hot electrons
in galaxy clusters without any simplifying assumptions.  We have
analytically reduced the collision term in the Boltzmann equation to
two dimensional integrals, and accurately solved the kinetic
equation. Our analytical expression for the collision integral is
exact, and can be used for any temperatures, $T_\gamma$, $T_e$, and
optical depth, $\tau$. The results are in good agreement with the
previous works for small temperature and optical depth, but
considerably different for large $\tau$. Our method can also be
applied in the case of an isotropic electron distribution function,
not necessarily in equilibrium.

\acknowledgments

We thank N.~Itoh, Y.~Kawana, S.~Kusano, and S.~Nozawa for sending us
their program for the data fits from ref.~\cite{ItohIV}.  A.D.~is
grateful to the Theory Division of CERN for the hospitality during the
period when this work was completed.  A.D., S.H.~and S.P.~are
supported by INFN.  In Munich, this work was partly supported by the
Deut\-sche For\-schungs\-ge\-mein\-schaft under grant No.\ SFB
375. The work of D.S.~is supported in part by INTAS grant 1A-1065.
S.P.~is supported by the European Commission under the TMR network
grant ERBFMRX-CT96-0090 and a Marie Curie fellowship under contract
HPMFCT-2000-00445.



\appendix
\section{Reduction of the collision integral}
\label{reduction}

In this appendix we list the expressions for the 7 terms in the
collision integral corresponding to the 7 terms in the matrix element
eq.~(\ref{matrix})
\beq
I_k (p_1) = \frac{\alpha^2}{2\pi p_1^2} \int_0^\infty~
\frac{p_2dp_2}{E_2}~\int_0^{{\cal S}}~dp_3~F~ J_k(p_1,p_2,p_3) \, .
\label{2-dimb}
\eeq
The functions $J_1, \ldots , J_7$ depend upon the following
combinations of the particle 3-momenta
\be
{\cal P}^2 \equiv (p_1+E_2)^2-m^2~, & &
{\cal Q}^2 \equiv (E_2-p_3)^2-m^2~, \nonumber \\
a = \max (|p_1-p_2|,|p_3-p_4|)~, & &b = \min (p_1+p_2,p_3+p_4)~,
\nonumber \\
c= \max (|p_1-p_4|,|p_2-p_3| )~, & & d= \min (p_1+p_4,p_2+p_3) \, .
\label{abcd}
\ee
In each case the value of $p_4$ is obtained from the condition of
energy conservation
\be
p_1 + \sqrt { p_2^2 +m^2} =p_3 + \sqrt { p_4^2 +m^2}~.
\label{e-cons}
\ee
The integration over $d^3p_4$ and the angles of ${\mathbf p}_2$ and
${\mathbf p}_3$ with $\delta$-function imposing energy-momentum
conservation is straightforward but rather lengthy, and we
present only the final results
\be
J_1 &=&  m^4~
\left [\frac{1}{{\cal P}^3}
\log \left| \frac{({\cal P}+b)({\cal P}-a)}{({\cal P}-b)({\cal P}+a)} \right|
+\frac{2 (b - a) 
\left(1 + a b/{\cal P}^2 \right)}{({\cal P}^2 - b^2)({\cal P}^2 - a^2)}
\right ]~, \\
J_2 &=&  - \frac{2 m^2}{p_1 p_3} (d - c) = -\frac{2 m^2}{p_1 p_3} (b - a)~,\\
J_4 &=& \frac{2 m^2}{{\cal P}}\log \left| \frac{({\cal P}+b)({\cal P}-a)}
{({\cal P}-b)({\cal P}+a)} \right|~, \\
J_6 &=& \frac{b-a}{2} \left(1  + 
\frac{(p_1^2 - p_2^2) ( p_3^2 - p_4^2)}{{\cal P}^2 a b}\right)  + 
\frac{m^2 p_1 p_3}{{\cal P}^3} 
\log \left| \frac{({\cal P}+b)({\cal P}-a)}{({\cal P}-b)({\cal P}+a)} \right|~.
\ee
The explicit form of the other three $J_i$ 
depends on the sign of ${\cal Q}^2$. If 
${\cal Q}^2 > 0$ then
\be
J_3 &=& m^4\left [\frac{1}{{\cal Q}^3}
\log \left | \frac{({\cal Q}+d)({\cal Q}-c)}{({\cal Q}-d)({\cal Q}+c)} \right|
+\frac{2 (d - c) 
\left(1 + c d/{\cal Q}^2 \right)}{({\cal Q}^2 - d^2)({\cal Q}^2 - c^2)}
\right ]~, \\
J_5 &=& \frac{2m^2}{{\cal Q}}
\log \left | \frac{({\cal Q}+d)({\cal Q}-c)}{({\cal Q}-d)({\cal Q}+c)} 
\right|~,\\
J_7 &=& \frac{d-c}{2}\left(1  + 
\frac{(p_1^2 - p_4^2) ( p_3^2 - p_2^2)}{{\cal Q}^2 c d}\right)  + 
\frac{m^2 p_1 p_3}{{\cal Q}^3} 
\log \left| \frac{({\cal Q}+d)({\cal Q}-c)}
{({\cal Q}-d)({\cal Q}+c)} \right| ~,
\ee
while for the case ${\cal Q}^2 < 0$, defining ${\cal W}^2=-{\cal Q}^2$ we find
\be
J_3 &=& 2m^4\left [\frac{1}{{\cal W}^3}
\arctan \left( \frac{(d-c){\cal W}}{{\cal W}^2+ c d}\right)
+\frac{(d - c) \left(1 + c d/{\cal Q}^2 \right)}
{({\cal Q}^2 - d^2)({\cal Q}^2 - c^2)}\right ]~, \\
J_5  &=& - \frac{4m^2}{{\cal W}}
\arctan \left( \frac{(d-c){\cal W}}{c d - {\cal Q}^2}\right) ~,\\
J_7 &=& \frac{d-c}{2} \left(1  + 
\frac{(p_1^2 - p_4^2) ( p_3^2 - p_2^2)}{{\cal Q}^2 c d}\right)  - 
\frac{2 m^2 p_1 p_3}{{\cal Q}^2{\cal W}} 
\arctan \left( \frac{(d-c){\cal W}}{c d - {\cal Q}^2}\right) \, .
\ee

\section{Integration region}
\label{integration}

In figs.~\ref{fig:regionsa} and \ref{fig:regionsb} we present the
integration region for the two-dimensional integral (\ref{2-dimb}),
where we took the particular case $p_1=1, m=50~p_1$ and measure momenta
$p_i$ in units of $p_1$.  The line $p_3={\cal S}=p_1+E_2-m$ is the
formal upper limit of integration over $p_3$, so $0 \le p_3 \le {\cal
S}$.  The integration region for $p_2$ is $0<p_2< \infty$, but in
fig.~\ref{fig:regionsb} we present it only for $p_2<m$.  Throughout
the paper we consider the case of non-relativistic electrons, $p_e
\la m$ but since we never expand in the electron mass, our analysis
is valid for the general (relativistic) case.

We can cut away most of the integration region for $p_3$ from the
condition $p_i \leq p_a + p_b +p_c$, which follows from the momentum
conservation, i.e. one momentum cannot be larger than the sum of all
the others.  For $p_i= p_1$ or $p_3$ this condition gives no
reduction, but in case of $p_i = p_2$ or $p_4$ we cut regions $E$ and
$F$ from the integration region\footnote{Note that region $E$
only exists if $p_1<m/2$.}  (see figs.~\ref{fig:regionsa} and
\ref{fig:regionsb}).  Thus only the regions $A$, $B$, $C$ and $D$
survive. To quantify this reduction let us note, that in the case of
CMB distortion in hot galaxy clusters one has $p_e \sim \sqrt{2 m T_e}
\sim 100~{\rm keV}$, while $p_\gamma \sim 10^{-4}~{\rm eV}$. In this
case formally $0<p_3<{\cal S} \sim 100~{\rm eV}$, while non-zero
contributions to the integrals in eq.~(\ref{2-dimb}) come only from
$p_3 \sim 10^{-4} ~{\rm eV}$.

Figure \ref{fig:regionsa} shows the integration region for relatively
small momenta. One can see that besides the main integration regions
$C$ and $D$ there are two small regions $A$ and $B$ which appear due
to the fact that for a small momentum $p_2$, the condition
$p_2=p_1+p_3+p_4$ is not operative, and the upper limit for $p_3$
becomes $p_3 = {\cal S}$.

The regions $A,B,C$ and $D$ are defined in such a way, that in each of
them our parameters $a$, $b$, $c$ and $d$ take definite values.  This
allows one to simplify the collision integral, $I_{coll}$, in each
region.  As we have seen in the previous section, there are generally two
kinds of contribution to the collision integral, like $J_1$ (which we
will call ${\cal P}$-kind) and like $J_3$ (which we will call ${\cal
Q}$-kind).  Fortunately the integration regions are identical for all
integrals, and only the limits $a,b,c$ and $d$ will differ for the two
kinds. For ${\cal Q}$-kind integrals we must distinguish between
positive and negative ${\cal Q}^2$ in the regions C and D (because
${\cal Q}^2$ is always negative in regions A and B).  Therefore we
introduce the regions $C_1$ and $D_1$ for ${\cal Q}^2<0$, and $C_2$ and $D_2$
regions for ${\cal Q}^2>0$.  We have also separated the contribution from the
${\cal Q}^2=0$ case, but this one-dimensional integral gives negligible
contribution to the two-dimensional collision integral.  Below we
define each of the regions in detail.

{\large \bf Region A.}
$$
0 < p_2 < p_1 ~~;~~ p_1 < p_3 <   p_1+E_2-m \, ,
$$
\centerline{and}
\begin{itemize}
\item If $p_1<m/2$ then,
$$
p_1 < p_2 < p_{change} ~~;~~ \frac{p_1(E_2+p_2)}{2 p_1-p_2+E_2} < p_3 <   p_1+E_2-m \, ,
$$
where $p_{change}=2 p_1 (m -p_1)/(m-2 p_1)$ is the point in 
which $p_2=p_1+p_3$ and $p_4=0$. 

\item If $p_1 \geq m/2$ then,
$$
p_1 < p_2 < \infty ~~;~~ \frac{p_1(E_2+p_2)}{2 p_1-p_2+E_2} < p_3 <   p_1+E_2-m \, ,
$$
and region $E$ does not exist.
\end{itemize}

In region $A$ we have
\[
a = p_3 - p_4 ~~;~~ b = p_3 + p_4 ~~;~~
c = p_1 - p_4 ~~;~~ d=p_1 + p_4 \, .
\]

\vskip1cm
{\large \bf Region B.}
$$
0 < p_2 < p_1 ~~;~~ \frac{p_1(E_2+p_2)}{2 p_1-p_2+E_2} < p_3 <   p_1 \, .
$$
In this region we have 
\[
a = p_1 - p_2 ~~;~~ b = p_1 + p_2  ~~ ; ~~
c = p_3 - p_2 ~~;~~ d=p_2 + p_3 \, .
\]

\vskip1cm
{\large \bf Region ${\bf C_1}$.}

In this region ${\cal Q}^2<0$. It is defined as 
$$
0 < p_2 < \min(p_1,p_{C_1}) ~~;~~  \frac{p_1(E_2-p_2)}{2 p_1+p_2+E_2}< p_3 
< \frac{p_1(E_2+p_2)}{2 p_1-p_2+E_2} \, ,
$$
\centerline{and}
\begin{itemize}
\item If $p_1 < m/2$ then,
$$
p_1 < p_2 < p_{C_1} ~~;~~  \frac{p_1(E_2-p_2)}{2 p_1+p_2+E_2} < p_3 < p_1 \, ,
$$
\item If $p_1 \geq m/2$ then,
$$
p_{C_1} < p_2 < p_1~~;~~  E_2-m < p_3 < \frac{p_1(E_2+p_2)}{2 p_1-p_2+E_2}\, ,
$$
\end{itemize}
\centerline{and}
$$
\max(p_1,p_{C_1}) < p_2 < p_{C_2} ~~;~~  E_2-m < p_3 <   p_1 \, ,
$$
where 
\be
p_{C_1}= \left\{ 
\begin{array}{cc}
\frac{(m-2 p_1)\sqrt{p_1(2 m + p_1)} - 2 p_1 (m - p_1)}{m - 4 p_1} & p_1 
\neq \frac{m}{4}\\
&\\
\frac{5}{12} m &  p_1=\frac{m}{4} \, , 
\end{array}
\right. 
\label{p_C1}
\ee
is the point, in which the line ${\cal Q}^2 = 0$ crosses the
line $p_4=p_1+p_2+p_3$ (see figures 
\ref{fig:regionsa} and \ref{fig:regionsb}).  
Let us just note the limiting cases for $p_{C_1}$
\be
p_1 \ll m && p_{C_1} \rightarrow \sqrt{2 p_1 m} \, ,
\nonumber \\ 
p_1 = \frac{m}{2} && p_{C_1} = p_1 =  \frac{m}{2} \, ,
\nonumber \\
p_1 \rightarrow \infty && p_{C_1} = \frac{3}{4} m \, .
\label{limits_p_C1}
\ee

The point in which the line ${\cal Q}^2 = 0$ crosses the line $p_3=p_1$ is 
\be
p_{C_2}= \sqrt{p_1(2 m + p_1)} \, .
\label{p_C2}
\ee

In region $C_1$ we have
\[
a = p_4 - p_3 ~~;~~ b = p_1 + p_2 ~~;~~
c = p_4 - p_1 ~~;~~ d=p_2 + p_3 \, .
\]

\vskip1cm
{\large \bf Region $\bf{C_2}$.}

In this region ${\cal Q}^2>0$. It is defined as 
$$
p_{C_1} < p_2 < p_{C_2} ~~;~~  \frac{p_1(E_2-p_2)}{2 p_1+p_2+E_2} 
< p_3 <  E_2 - m \, ,
$$
and 
$$
p_{C_2} < p_2 < \infty ~~;~~  \frac{p_1(E_2-p_2)}{2 p_1+p_2+E_2} < p_3 < p_1 
\, .
$$
The limits $a$, $b$, $c$ and $d$  are the same as in region  $C_1$. 

\vskip1cm
{\large \bf Region ${\bf D_1}$.}

In this region ${\cal Q}^2<0$. It is defined as 
$$
p_1 < p_2 < p_{C_2} ~~;~~  
p_1< p_3 < \frac{p_1(E_2+p_2)}{2 p_1-p_2+E_2} \, ,
$$
and 
$$
p_{C_2} < p_2 < p_D ~~;~~  
E_2 - m< p_3 < \frac{p_1(E_2+p_2)}{2 p_1-p_2+E_2} \, ,
$$
where $p_D$ is the point where the ${\cal Q}^2=0$ line crosses the 
line $p_2=p_1+p_3+p_4$. This point is defined as 
\be
p_{D}= \left\{ 
\begin{array}{cc}
\frac{(m-2 p_1)\sqrt{p_1(2 m + p_1)} + 2 p_1 (m - p_1)}{m - 4 p_1} & p_1 < \frac{m}{4}\\
&\\
3m/(4\epsilon) &  p_1 = \frac{m}{4}- m \epsilon, ~~~\epsilon \ll 1\\
&\\
\infty & p_1 \ge m/4  \, .\\
\end{array} 
\right.
\label{p_D}
\ee

In region $D_1$  we have 
\[
a = p_2 - p_1 ~~;~~ b = p_3 + p_4 ~~ ; ~~
c = p_2 - p_3 ~~;~~ d=p_1 + p_4 \, .
\]

{\large \bf Region ${\bf D_2}$.}

In this region ${\cal Q}^2>0$. It is defined as 
$$
p_{C_2} < p_2 < p_D ~~;~~  
p_1< p_3 < E_2 - m \, ,
$$
and if $p_D$ is finite
$$
p_D < p_2 < \infty ~~;~~  
p_1< p_3 < \frac{p_1(E_2+p_2)}{2 p_1-p_2+E_2} \, .
$$
The limits $a$, $b$, $c$ and $d$ are the same as in $D_1$.


\clearpage



\figcaption[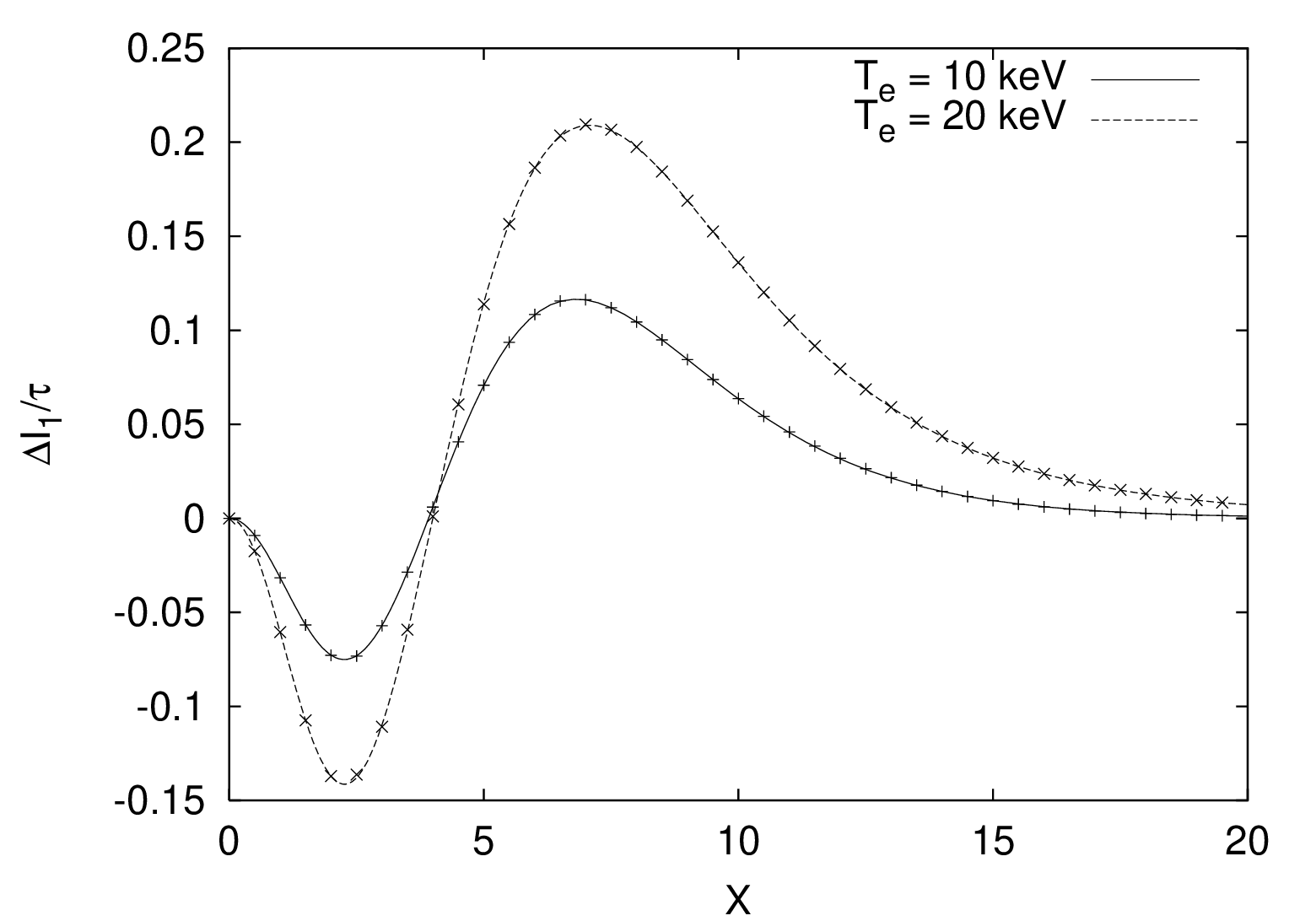]{First-order contribution to the spectral
distortion of photons for $T_e=10$ keV and $T_e=20$ keV as a function
of the dimensionless photon frequency $X$.  Our numerical results are
presented by the solid and dashed lines, whereas the data points
correspond to the analytical fit from \citet{ItohIV}. \label{fig:I1}}

\figcaption[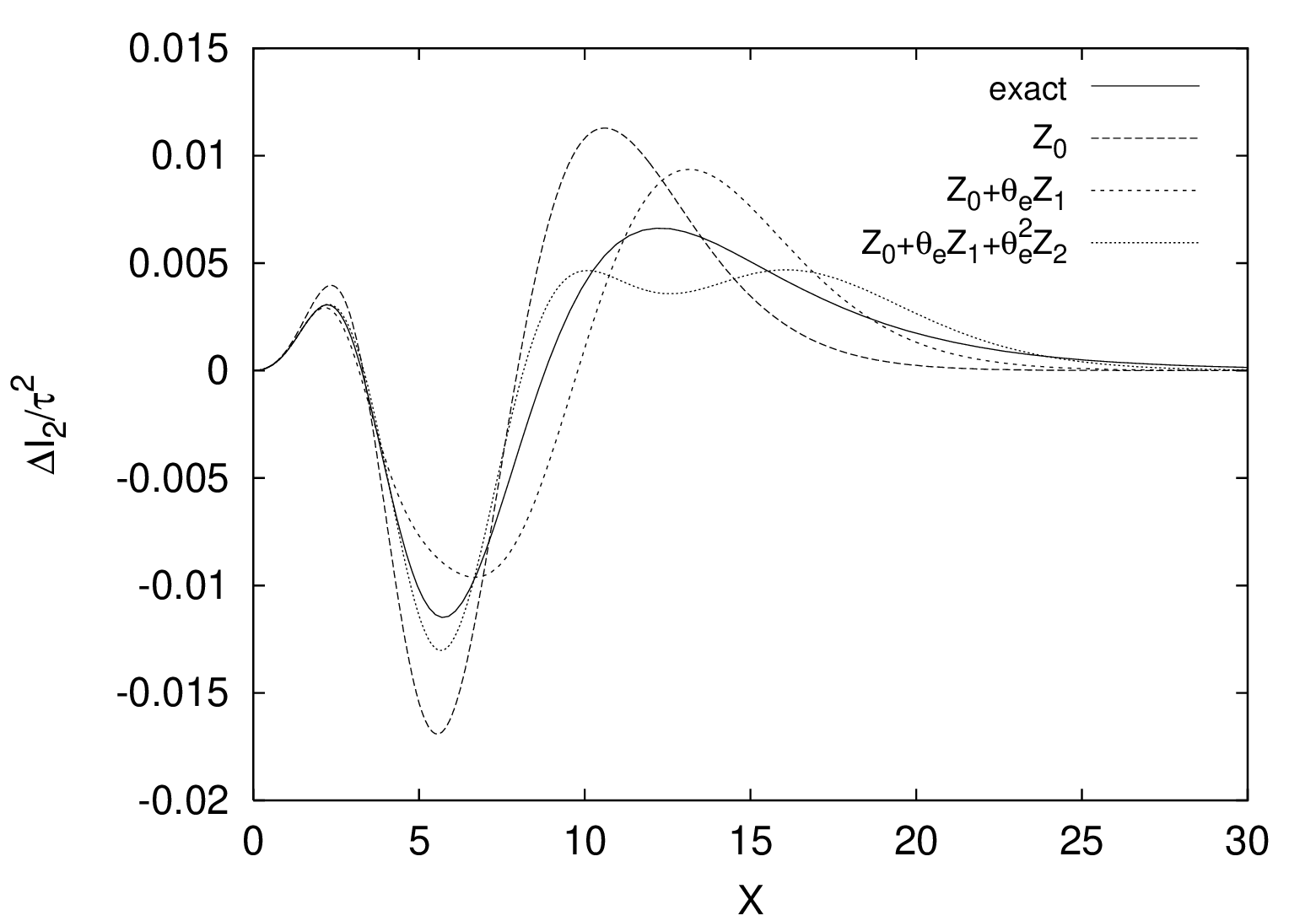]{Second-order contribution to the spectral
distortion of photons for $T_e=10$ keV as a function of the
dimensionless photon frequency $X$.  Our numerical result are
presented by the solid line, whereas the dashed, short dashed and
dotted lines correspond to the series expansion from
\citet{ItohV}. \label{fig:I2}}

\figcaption[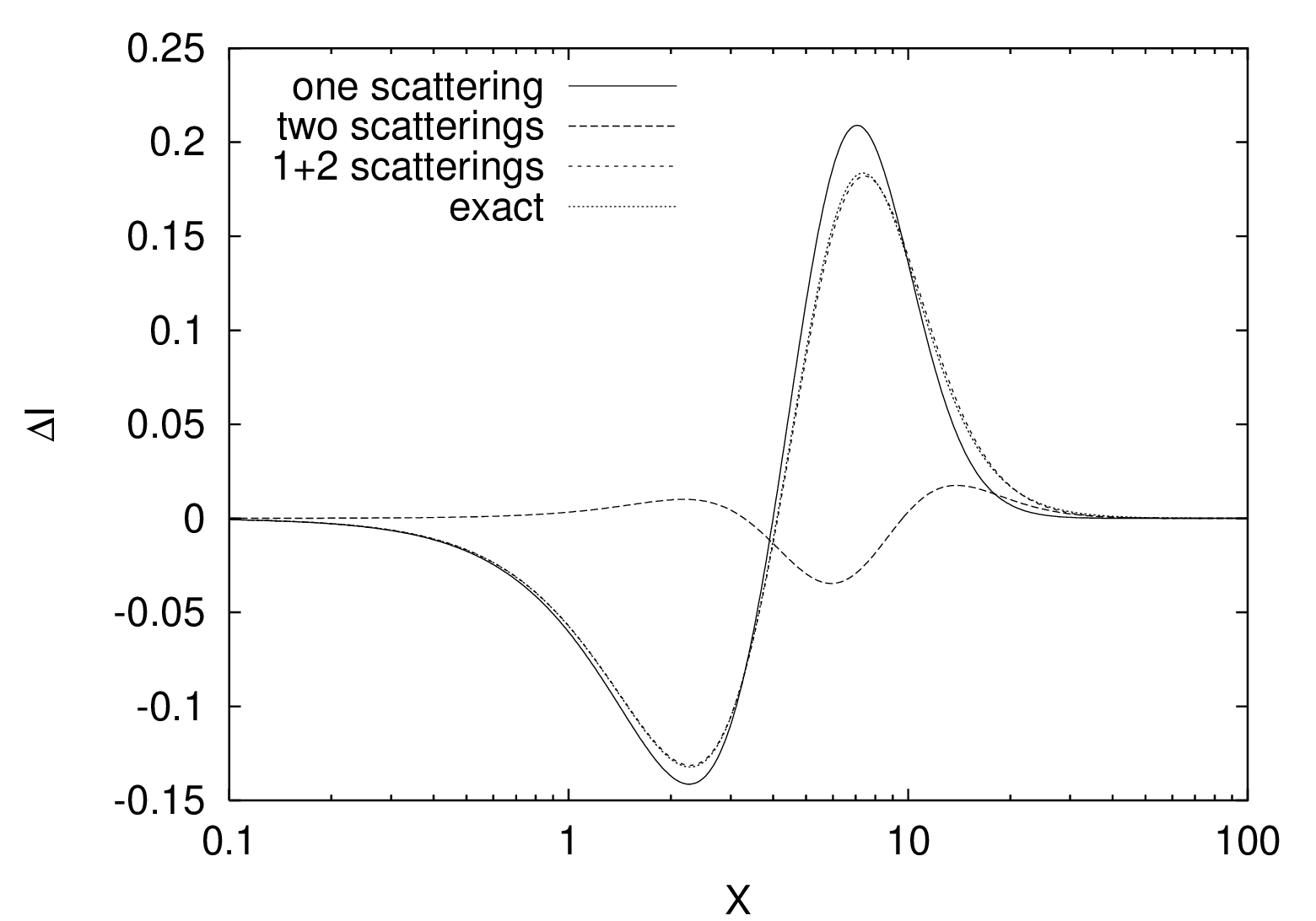]{Spectral distortion of photons for optical depth
$\tau=1$ and $T_e=20$ keV as a function of the dimensionless photon
frequency $X$. \label{fig:T20y1}}

\figcaption[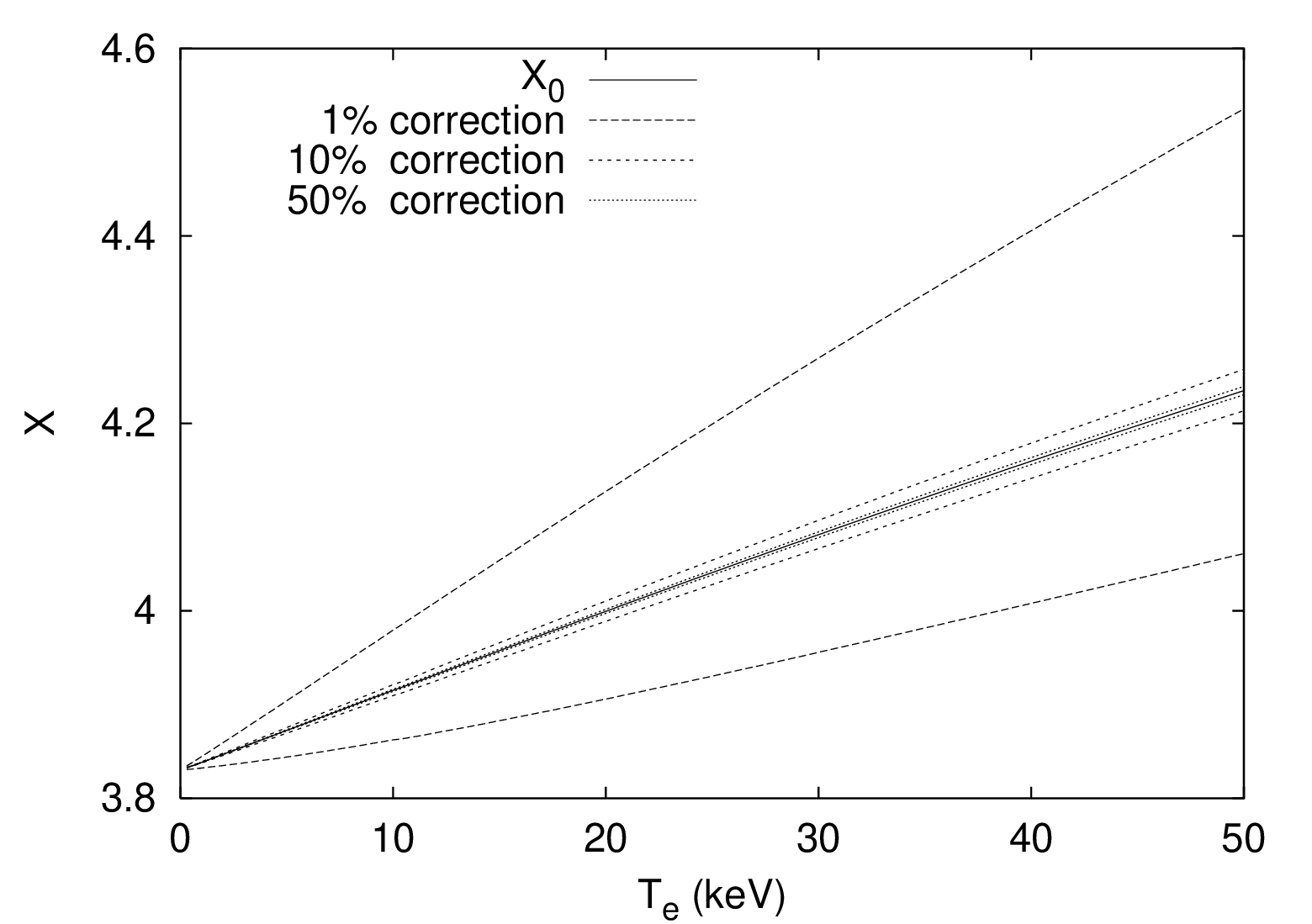]{Crossover frequency $X_0$ as a function of the
electron temperature $T_e$ for $\tau = 0.01$ and the regions, where
double photon scattering gives a contribution to the spectral
distortion larger than 1\% (long dashed lines), 10\% (short dashed
lines), and 50\% (dotted lines) respectively. \label{fig:corr_Te}}

\figcaption[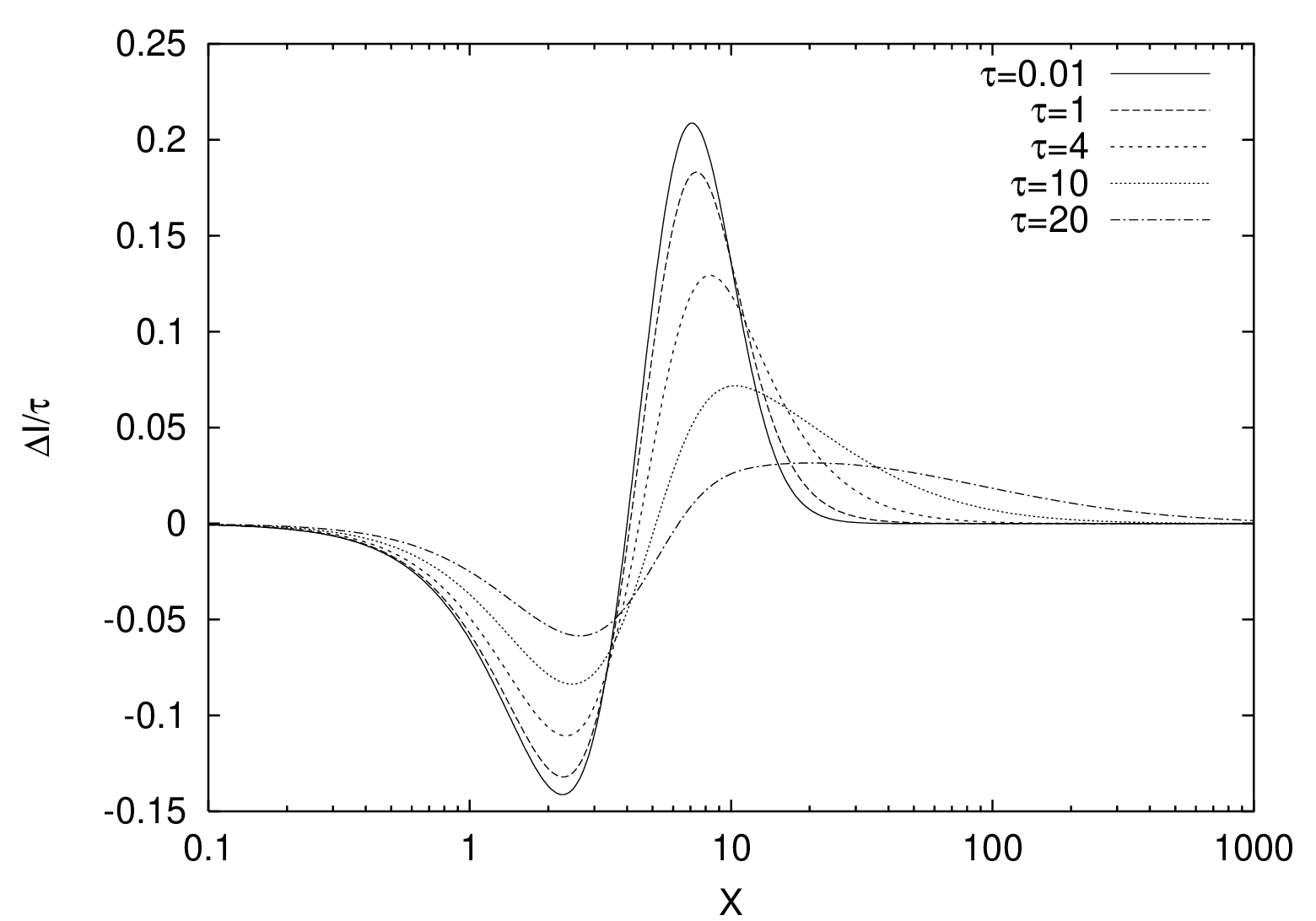]{Spectral distortion of photons for  $T_e=20$ keV
as a function of the dimensionless photon frequency $X$ for several
optical depths up to $\tau=20$. \label{fig:T20y20}}

\figcaption[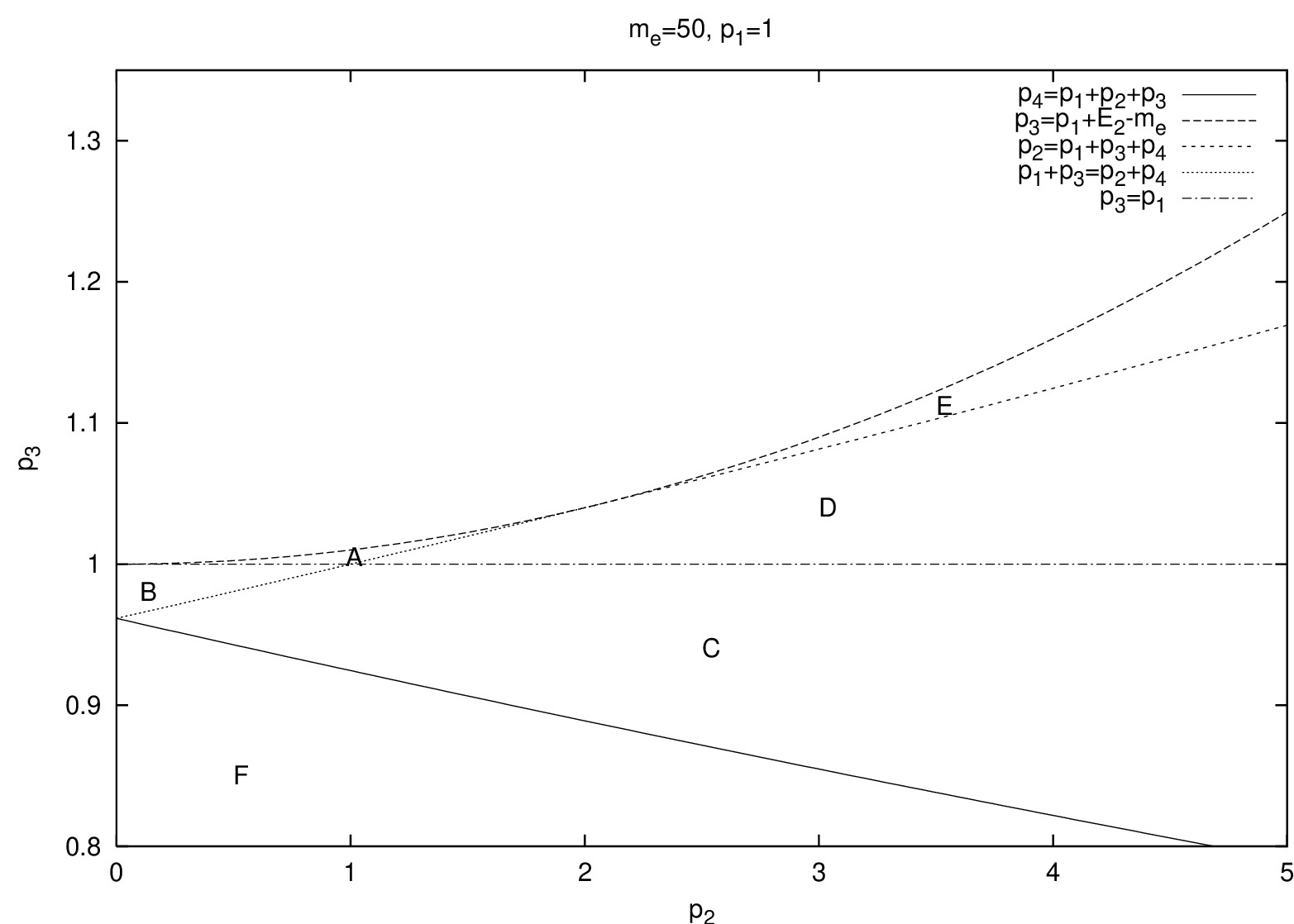]{Region of integration of two-dimensional integral
for small momenta. \label{fig:regionsa}}

\figcaption[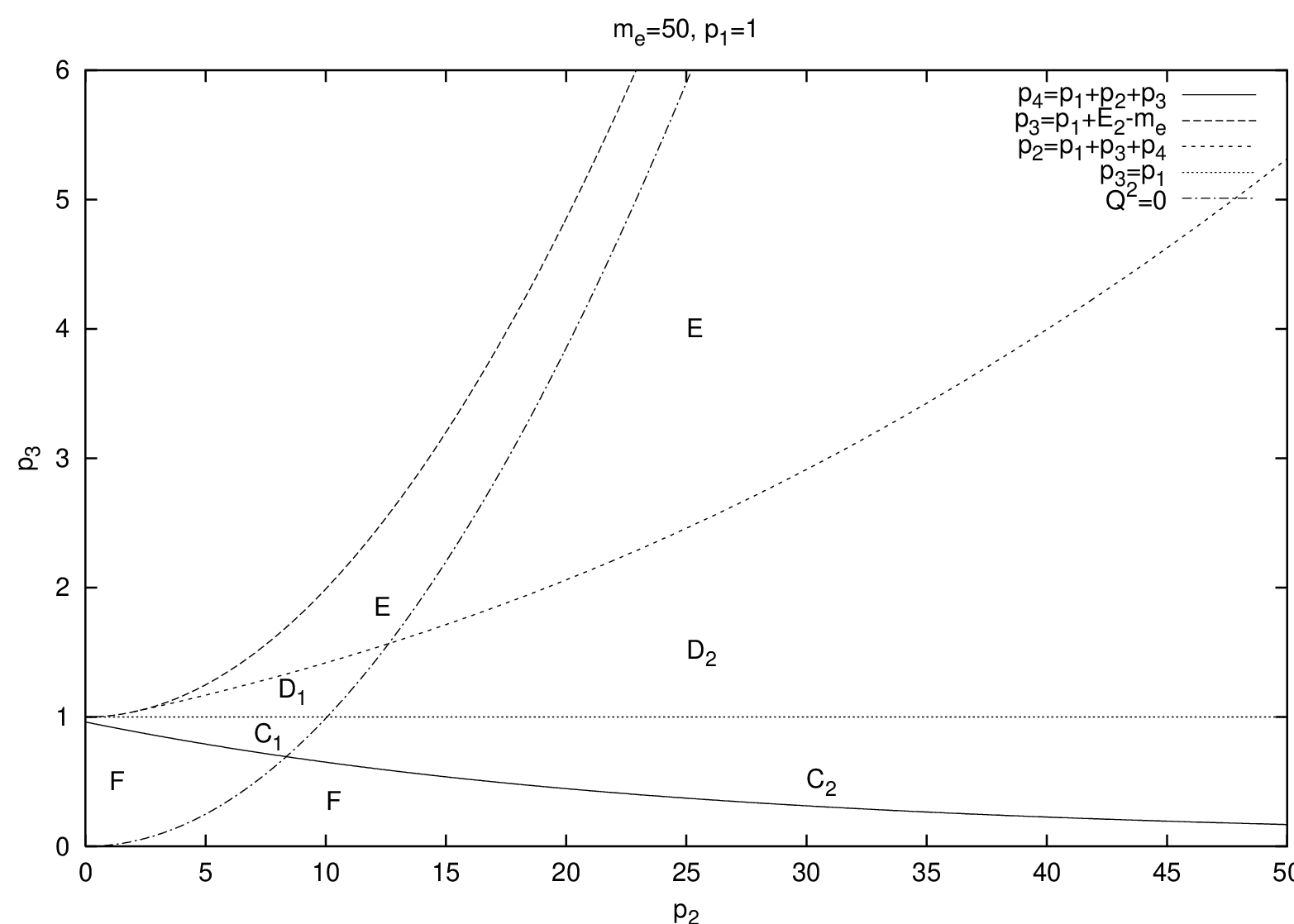]{Region of integration of two-dimensional integral.
\label{fig:regionsb}}

\end{document}